\documentclass[twocolumn,aps,prl,amsmath,superscriptaddress,longbibliography,notitlepage]{revtex4}
\usepackage{times}
\usepackage{txfonts}
\usepackage{amssymb}
\usepackage{mathrsfs}
\usepackage{graphicx}
\usepackage{float}
\usepackage[caption=false]{subfig}
\usepackage[colorlinks,bookmarks=true,citecolor=blue,linkcolor=blue,urlcolor=blue]{hyperref}
\usepackage{verbatim}
\usepackage{epstopdf}
\usepackage[normalem]{ulem}
\usepackage{verbatim}
\usepackage{xcolor}
\usepackage{bm}
\usepackage{physics}
\usepackage{ulem}

\renewcommand{\Im}{\operatorname{Im}}
\newcommand{\clr}{\color{red!75!black}}

\begin{document}
\title{Dynamical Poles in Non-Hermitian Impurity Scattering}
\author{Ao Yang}
\email{yangao@iphy.ac.cn}
\affiliation{Beijing National Laboratory for Condensed Matter Physics, and Institute of Physics, Chinese Academy of Sciences, Beijing 100190, China}
\affiliation{University of Chinese Academy of Sciences, Beijing 100049, China}

\author{Kai Zhang}
\affiliation{Department of Physics, University of Michigan Ann Arbor, Ann Arbor, Michigan, 48109, United States}

\author{Chen Fang}
\affiliation{Beijing National Laboratory for Condensed Matter Physics, and Institute of Physics, Chinese Academy of Sciences, Beijing 100190, China}
\affiliation{Kavli Institute for Theoretical Sciences, Chinese Academy of Sciences, Beijing 100190, China}

\begin{abstract}
In Hermitian impurity scattering, each isolated late-time exponential is the fingerprint of a bound state.
We show that this correspondence breaks down in non-Hermitian bands.
For a single impurity in a non-Hermitian lattice, the late-time signal is controlled by isolated complex frequencies selected by the analytic continuation of the Green's function relevant to real-time dynamics, which we term dynamical poles (DPs).
DPs need not coincide with static bound states: one may appear without any bound-state counterpart, while a static bound state may be dynamically invisible.
The remainder of the signal is an incoherent background set by complex continuum edges.
Our results establish that the real-time analytic structure of the Green's function, not the static eigenvalue problem alone, organizes non-Hermitian impurity scattering.

\end{abstract}

\maketitle

\emph{{\clr Introduction.}---}~
A cornerstone of Hermitian scattering theory is the direct link between the static spectrum and the long-time scattering dynamics~\cite{Philip2012,Weinberg1995,Economou2006}.
The mechanism behind this link is the analytic structure of the Green's function: each bound state corresponds to an isolated pole, whose residue yields a discrete exponential in the late-time signal, while the continuous spectrum gives rise to a branch cut whose contribution is subleading, i.e., decays algebraically.
Every late-time discrete exponential is thus the fingerprint of a static bound state, and conversely, every bound state can be tracked dynamically.
This fundamental correspondence underlies 
a broad range of scattering and transport probes, from resonant optical response in photonic lattices and microcavities~\cite{Huber2025,Plotnik2011,Foreman2015} to impurity problems in cold-atom systems~\cite{Massignan2014,Cetina2016,Knap2012} and quantum-coherent transport in mesoscopic quantum dots ~\cite{Alhassid2000,Guttinger2012,Franceschi2010}.

This correspondence, however, does not necessarily persist in non-Hermitian systems. 
In the above experimental platforms, radiative loss, optical pumping, or particle decay naturally introduce non-Hermiticity~\cite{Rotter_2009,Diehl2011,Simon2015,Jean1992,Regensburger2012,Gao2015_Nature,FengLiang2017,Ganainy2018,Miri2019,YangLan2019,Kozii2024,ShenHT2018_PRL,FuLiang2020_PRL,SongF2019_PRL,Ashida2020}.
A non-Hermitian Hamiltonian generically has complex eigenvalues and generates non-unitary dynamics, so that amplitudes may grow or decay and the long-time signal need not decompose into contributions dictated by the static eigenstates.
Much recent work has explored non-Hermitian band theory~\cite{Lee2016PRL,luisPRB2018,Yao2018,Kunst2018_PRL,WangZhong2018,Murakami2019_PRL,ChingHua2019,LeeCH2019_PRL,LonghiPRR2019,Kai2020,Okuma2020_PRL,Slager2020PRL,Zhesen2020_aGBZ,Zhesen2020_SE,XuePeng2020,Ghatak2020,Thomale2020,LiLH2020_NC,Kawabata2020_Symplectic,Wanjura2020_NC,XueWT2021_PRB,LiLH2021_NC,Kai2022NC,ZhangDDS2022,Longhi2022PRL} and identified the counterparts of impurity bound states and scattering states in this setting~\cite{Fang2023,Yang2024,Yang2025}.
However, whether these static eigenstates govern the scattering dynamics in the same way as in the Hermitian case has not been established.
Indeed, even in homogeneous non-Hermitian lattices, long-time dynamics can be governed by saddle points away from the eigenvalue spectrum~\cite{LonghiPRR2019,Fang2025,Xue2025}, suggesting that the static spectrum alone may not determine the asymptotic signal.
The central question thus arises: in a non-Hermitian band with impurities, what actually determines the late-time signal?

In this Letter, we answer this question by showing that the relevant discrete frequencies are not the static bound-state energies, but poles of the analytically continued Green's function, which we call dynamical poles (DPs).
Each DP contributes a discrete exponential to the late-time signal, while the remainder is an incoherent background set by complex branch points that serve as non-Hermitian continuum edges.
Because DPs arise from the analytically continued Green's function rather than the static eigenvalue problem, the static-dynamical correspondence is generically broken in two complementary ways: a DP may exist without any bound-state counterpart, giving rise to a late-time exponential that has no spectral origin; conversely, a static bound state may be dynamically invisible, contributing to the eigenvalue spectrum but leaving no trace in the time-domain signal.
We call such states dynamically dark. 
This two-sided mismatch has direct physical consequences: a spectroscopic measurement that resolves the eigenvalue spectrum and a time-resolved measurement that tracks the late-time response will, in general, report different sets of discrete frequencies.
This framework applies naturally to various matrix elements~\cite{SupMat} and to open boundary conditions (OBC). As a consistency check, it also reproduces known results in Hermitian systems~\cite{Madrid2002,Correia2025,Wolkanowski2013,Yapa2023} and non-Hermitian homogeneous limits~\cite{LonghiPRR2019,Xue2022,Hu2024,Fang2025,Xue2025}. 

\emph{{\clr Dynamical poles beyond static bound states.}---}~
We first illustrate the mechanism in a one-dimensional Hatano-Nelson chain under periodic boundary conditions (PBC), $\hat H=\hat H_0+\hat V$, with a single onsite impurity $\hat V=\lambda\ket{0}\bra{0}$. 
The clean Hamiltonian has dispersion $E_k=e^{ik}+Je^{-ik}-i\kappa$ with $J\in(0,1)$ and $\kappa>0$, and we study the momentum-resolved amplitude
$M_{k'k}(t):=\bra{k'}e^{-i\hat H t}\ket{k}$. 
The uniform loss term $-i\kappa \hat I$ shifts the spectrum below the real axis and only contributes an overall decay factor $e^{-\kappa t}$.

\begin{figure}[t]
    \centering
	    \includegraphics[width=1\linewidth]{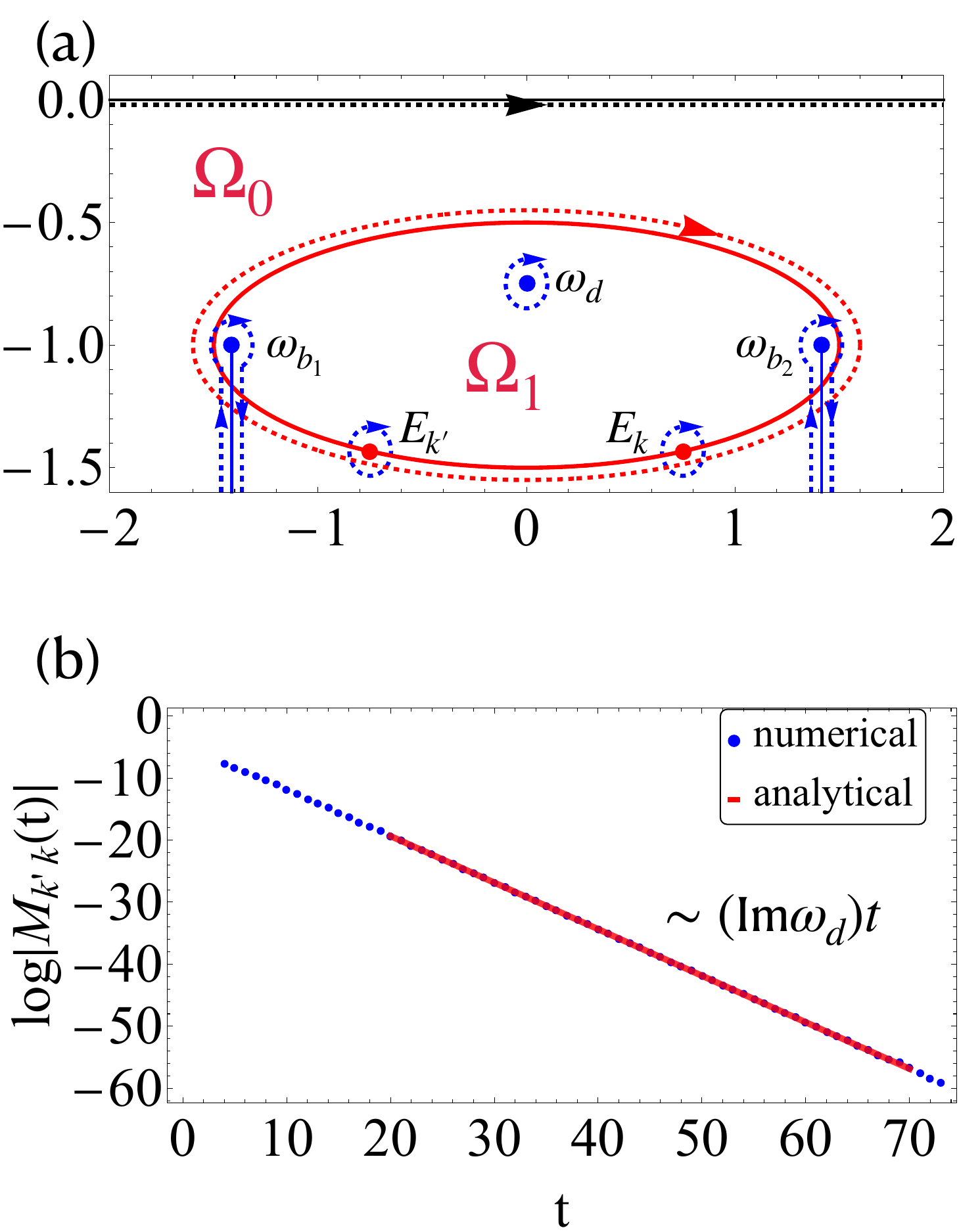}
	    \caption{
		(a) Red solid ellipse: spectrum of the perturbed HN model. Parameters: $J=1/2$, $\kappa=1$, $\lambda=\tilde{g}_0(\omega_d)^{-1}\approx 1.436i$, and $\omega_d=-0.75i$.
		After analytic continuation to $\tilde{g}_0$, the integral decomposes into three parts: (i) dashed blue circle around $\omega_d$ (DP); (ii) dashed blue circle around $E_k$ (initial state) and $E_{k'}$ (final state); (iii) dashed circles around $\omega_{b_{1,2}}$ and the attached cut lines (branch points). (b) $\log|M_{k'k}(t)|$ for initial state $k=-\pi/3$ and $k'=-2\pi/3$. The late-time signal is dominated by the DP at $\omega_d$.
		}
    \label{fig:f2}
\end{figure}

This model already provides a sharp counterexample to the static-spectrum intuition. 
For a single impurity, the $T$ matrix is
$\hat T(\omega)=\lambda\ket{0}\bra{0}/[1-\lambda g_0(\omega)]$,
where
$g_0(\omega)=\bra{0}(\omega-\hat H_0)^{-1}\ket{0}$,
so static impurity poles are determined by $1-\lambda g_0(\omega)=0$. 
However, the PBC spectral loop encloses a point gap inside which $g_0(\omega)$ vanishes identically [Fig.~\ref{fig:f2}(a)]~\cite{SupMat,Fang2023}.
Hence no static impurity pole exists there, even though Fig.~\ref{fig:f2}(b) exhibits a clean late-time exponential in $M_{k'k}(t)$.

The resolution lies in the frequency-domain representation of $M_{k'k}(t)$ (see the SM~\cite{SupMat}),
\begin{widetext}
\begin{equation}
M_{k'k}(t)
=
\delta_{k'k}e^{-iE_k t}
-\braket{k'}{0}\braket{0}{k}\,
\frac{\lambda}{2\pi i}\int_{-\infty}^{\infty}d\omega\;
\frac{e^{-i\omega t}}{(\omega-E_k)(\omega-E_{k'})(1-\lambda g_0(\omega))}.
\label{eq:toy_momentum_integral_v5}
\end{equation}
\end{widetext}

The integral depends only on the values of $g_0(\omega)$ on the real axis.
For $t>0$, one closes the contour into $\Im\omega<0$ to extract the long-time behavior; different analytic continuations of $g_0$ off the real axis all yield the same result but organize it into different contributions.
For example, deforming the contour onto the PBC spectral loop [red dashed contour in Fig.~\ref{fig:f2}(a)] decomposes the signal as an integral over non-Hermitian scattering states.
This representation is exact but mixes all frequencies together. 
In contrast to the Hermitian case, where only the on-shell scattering state contributes to the discrete late-time signal, here all scattering states contribute, and the discrete exponentials that dominate at late times are not directly visible.

A more transparent decomposition follows from the straightforward analytic continuation, $\tilde g_0(\omega)=[(\omega+i\kappa)^2-4J]^{-1/2}$, defined on $\mathbb C\setminus(\Gamma_1\cup\Gamma_2)$ with branch points $\omega_{b_{1,2}}=\pm2\sqrt{J}-i\kappa$ and downward cuts $\Gamma_{1,2}$ [blue dashed contour in Fig.~\ref{fig:f2}(a)].
Unlike the BZ-defined $g_0$, which vanishes inside the point gap, $\tilde g_0$ is generically nonzero there, so the continued denominator $1-\lambda\tilde g_0(\omega)$ can have zeros.
These are the dynamical poles(DPs), $\tilde S=\{\omega_d:1-\lambda \tilde g_0(\omega_d)=0\}$, invisible to the static Green's function but revealed by the continuation.

With this continuation, the long-time signal separates into three contributions,
\begin{equation}
\begin{split}
M_{k'k}(t)
\sim
\sum_{\omega_d\in \tilde{S}} C_{\omega_d}^{k'k}e^{-i\omega_d t}
+
\sum_{\omega\in\{E_k,E_{k'}\}} C_{\omega}^{k'k}e^{-i\omega t}
+
\sum_{\omega_b\in B} I_{\omega_b}^{k'k}(t).
\end{split}
\label{eq:toy_momentum_decomp_v2}
\end{equation}
The first term is the discrete contribution from DPs, the second is a kinematic contribution fixed by the incident and detected plane-wave states, and the third comes from branch points that play the role of continuum edges.
In one dimension,
$I_{\omega_b}^{k'k}(t)\sim t^{-3/2}e^{-i\omega_b t}$;
the corresponding prefactors, as well as analogous formulas for $\langle 0|e^{-i\hat H t}|k\rangle$ and $\langle 0|e^{-i\hat H t}|0\rangle$, are given in the SM~\cite{SupMat}.

This decomposition resolves the mismatch identified above.
At long times, the dominant contribution comes from the singularity with the largest $\Im\,\omega$.
A DP can therefore generate the leading late-time exponential even when no static impurity pole exists in the original Green's function.
This is precisely the behavior seen in Fig.~\ref{fig:f2}(b).

\emph{{\clr General mechanism: complex continuum edges and dynamical poles.}---}~
The Hatano-Nelson example reveals a general picture. The long-time signal is governed by the analytic continuation $\tilde g_0$ of the local Green's function, whose singularity structure can differ markedly from that of the BZ-defined $g_0$. Once the trivial kinematic contributions from the incident and detected states are set aside, the signal is controlled entirely by two features of $\tilde g_0$: dynamical poles, determined by $1-\lambda\tilde g_0(\omega)=0$, and branch points, which play the role of non-Hermitian continuum edges, as we show below.

In the Hatano-Nelson model, the explicit form of $\tilde g_0$ could be read off directly. For a general lattice Hamiltonian, however, the continuation must be constructed systematically. The central problem is to build $\tilde g_0(\omega)$ and identify where it becomes nonanalytic.

For a general one-dimensional lattice with Bloch Hamiltonian
$h(z\equiv e^{ik})=\bra{k}\hat H_0\ket{k}=\sum_{n=-N}^{M} t_n z^n$,
the local Green's function has the standard residue form
\begin{equation}
\begin{split}
g_0(\omega)&=\oint_{|z|=1}\frac{dz}{2\pi i\,z}\,\frac{1}{\omega-h(z)}\\
&=-\sum_{|z_j(\omega)|<1}\frac{1}{z_j(\omega)h'(z_j(\omega))},
\qquad (\Im\omega\ge 0),
\end{split}
\label{eq:g0general_ge_v2b}
\end{equation}
where $z_j(\omega)$ are the roots of $\omega-h(z)=0$. 
For $\Im\omega \gg 0$, all roots are unambiguously inside or outside the unit circle.
Precisely, $N$ roots lie inside $|z|=1$ and $M$ roots outside,
and the residue sum is well defined. 
The continuation problem is thus reduced to following the $N$ root branches that lie inside $|z|=1$ at such a reference point and analytically continuing them into $\Im\omega<0$.

Nonanalyticities arise when two roots coalesce, i.e., when $\omega-h(z)=0$ develops a multiple root at $(\omega_b,z_b)$ with $h'(z_b)=0$. 
However, not every such collision produces a branch point of $\tilde g_0$. 
Because $\tilde g_0$ is a sum over the selected root branches, a collision that merely permutes roots within the selected set leaves $\tilde g_0$ unchanged. 
A true branch point appears only when analytic continuation around $\omega_b$ forces an exchange between a selected inside branch and an unselected outside branch. 
We denote by $B$ the subset of multiple-root energies where this inside--outside exchange occurs. 
Note that during the continuation the roots may cross $|z|=1$, but their inside/outside identity is fixed by the classification at the reference point, not by their current position.

\begin{figure}[t]
    \centering
    \includegraphics[width=1\linewidth]{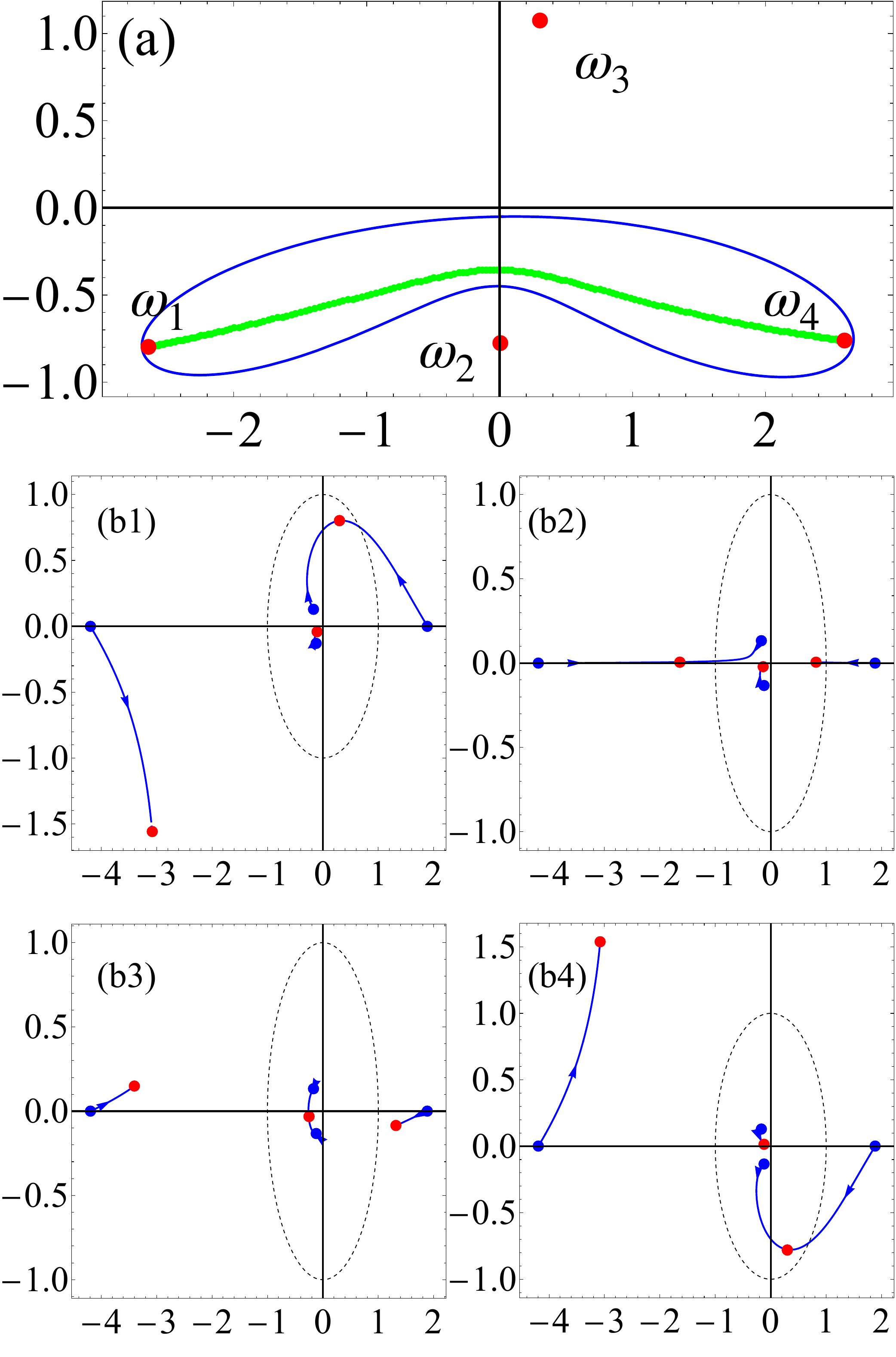}
    \caption{
	Illustration of the branch-point criterion for $\tilde{g}_0(\omega)$. The model is chosen as $h(z)=-0.06i + (0.02 -0.15i)z^{-2} -1.1 i z^{-1}+1.3i z+ 0.5iz^2$. (a) Blue curve: PBC spectrum; green curve: OBC spectrum; red points: four multiple-root points of $z(\omega)$, denoted by $\omega_j$, $j=1,2,3,4$. (b$j$): Root trajectories from $\omega_{\rm ref}=2i$ to $\omega_j$. The four blue points in each panel correspond to the four roots at $\omega_{\rm ref}$, and the red points correspond to the roots at $\omega_j$. Root collisions at $\omega_1,\omega_2,\omega_4$ mix branches that are initially inside and outside the unit circle (dashed black circle), while $\omega_3$ does not. Thus $B=\{\omega_1,\omega_2,\omega_4\}$.  
	}
    \label{fig:f3}
\end{figure}

Fig.~\ref{fig:f3} illustrates this criterion in a model with next-nearest-neighbor hopping. Among the four multiple-root energies shown there, $\omega_{1,2,4}$ mix inside and outside branches and therefore belong to $B$, whereas $\omega_3$ does not.

Once a branch point $\omega_b\in B$ is identified, its local form is universal because it comes from a generic quadratic root collision,
\begin{equation}
\omega-h(z)\simeq (\omega-\omega_b)-\frac{h''(z_b)}{2}(z-z_b)^2.
\label{eq:local_collision_ge_v4}
\end{equation}
Hence $z_\pm-z_b\propto\sqrt{\omega-\omega_b}$. Since $h'(z_b)=0$ at the collision, $h'(z_\pm)\sim h''(z_b)(z_\pm-z_b)\propto\sqrt{\omega-\omega_b}$, and the residue of each colliding root in Eq.~\eqref{eq:g0general_ge_v2b} diverges as $(\omega-\omega_b)^{-1/2}$.
The inside--outside exchange at $\omega_b\in B$ retains exactly one of the two colliding branches, thus such divergence survives in $\tilde g_0$:
$\tilde g_0(\omega)\sim (\omega-\omega_b)^{-1/2}$.
The same local analysis in higher dimensions shows that the nonanalytic part of $\tilde g_0(\omega)$ behaves as $(\omega-\omega_b)^{d/2-1}\log(\omega-\omega_b)$ for even $d$, and $(\omega-\omega_b)^{d/2-1}$ for odd $d$.
In every case, the singularity has the same form as the Hermitian band edge in the corresponding dimension~\cite{Joyce2003}, and it produces the same type of dynamical contribution: an exponential multiplied by a dimension-dependent algebraic decay prefactor, possibly with logarithmic corrections. For example, in one dimension, $I_{\omega_b}(t)\sim t^{-3/2}e^{-i\omega_b t}$.
The key difference is that $\omega_b$ is now generically complex, so the exponential carries an intrinsic decay rate.
These branch points therefore serve as the non-Hermitian generalization of continuum edges. The higher-dimensional analysis, cut-integral formulas, and explicit prefactors are given in the SM~\cite{SupMat}.

The analogy extends to the existence of discrete states.
In Hermitian systems with $d\le 2$, any arbitrarily weak impurity creates a bound state near the band edge, because the Green's function diverges there.
The same mechanism applies, with the band edge replaced by the branch point and the bound state by the DP, since $\tilde g_0$ diverges at any $\omega_b\in B$ for $d\le 2$ and the equation $1-\lambda\tilde g_0(\omega)=0$ has solutions for arbitrarily small values of $\lambda$.
For $d\ge 3$, $\tilde g_0$ remains finite at $\omega_b$ and a threshold impurity strength is required, again paralleling the Hermitian case.

\emph{{\clr Static-dynamical mismatch and dark bound states.}---}~
The static $g_0$ and its continuation $\tilde g_0$ agree on the upper half-plane and real axis, but need not coincide in the lower half-plane, where the long-time asymptotics is determined.
The branch cuts of $g_0$ (i.e., the spectrum) and of $\tilde g_0$ divide the $\omega$-plane into several connected regions.
By the uniqueness theorem for analytic continuation, $g_0=\tilde g_0$ on the region connected to the upper half-plane, but they can differ on all others.

We define the mismatch domain
$D:=\{\omega\in \Im\omega<0:\ g_0(\omega)\neq \tilde g_0(\omega)\}$.
In the Hatano-Nelson model, $D$ consists of the point-gap interior together with the region between the two branch cuts.

This characterization has two immediate consequences.
First, $1-\lambda\tilde g_0(\omega)=0$ may have a solution in $D$ where $1-\lambda g_0(\omega)=0$ does not; then a DP appears dynamically without any bound-state counterpart.
Second, a bound state at $\omega_0\in D$ solving $1-\lambda g_0(\omega_0)=0$ need not solve $1-\lambda\tilde g_0(\omega_0)=0$; then the state is present spectrally but contributes no pole term in the time-domain contour evaluation.
In this sense, bound states inside $D$ are dynamically dark.

\emph{{\clr OBC case.}---}~
The mismatch-domain framework is not a peculiarity of PBC.
For a bulk impurity site far from the boundaries, the local Green's function $g^{(\mathrm{OBC})}_0(\omega)=\langle 0|(\omega-\hat H^{(\mathrm{OBC})}_0)^{-1}|0\rangle$ coincides with $g_0(\omega)$ on the real axis, since the BZ contour can be continuously deformed to the GBZ contour without crossing singularities.
The only change is that the OBC spectrum replaces the PBC spectrum in the geometric criterion above: the OBC spectrum and the branch cuts of $\tilde g_0$ divide the $\omega$-plane into connected regions, and $D$ is redefined accordingly.

\begin{figure}[t]
    \centering
    \includegraphics[width=1\linewidth]{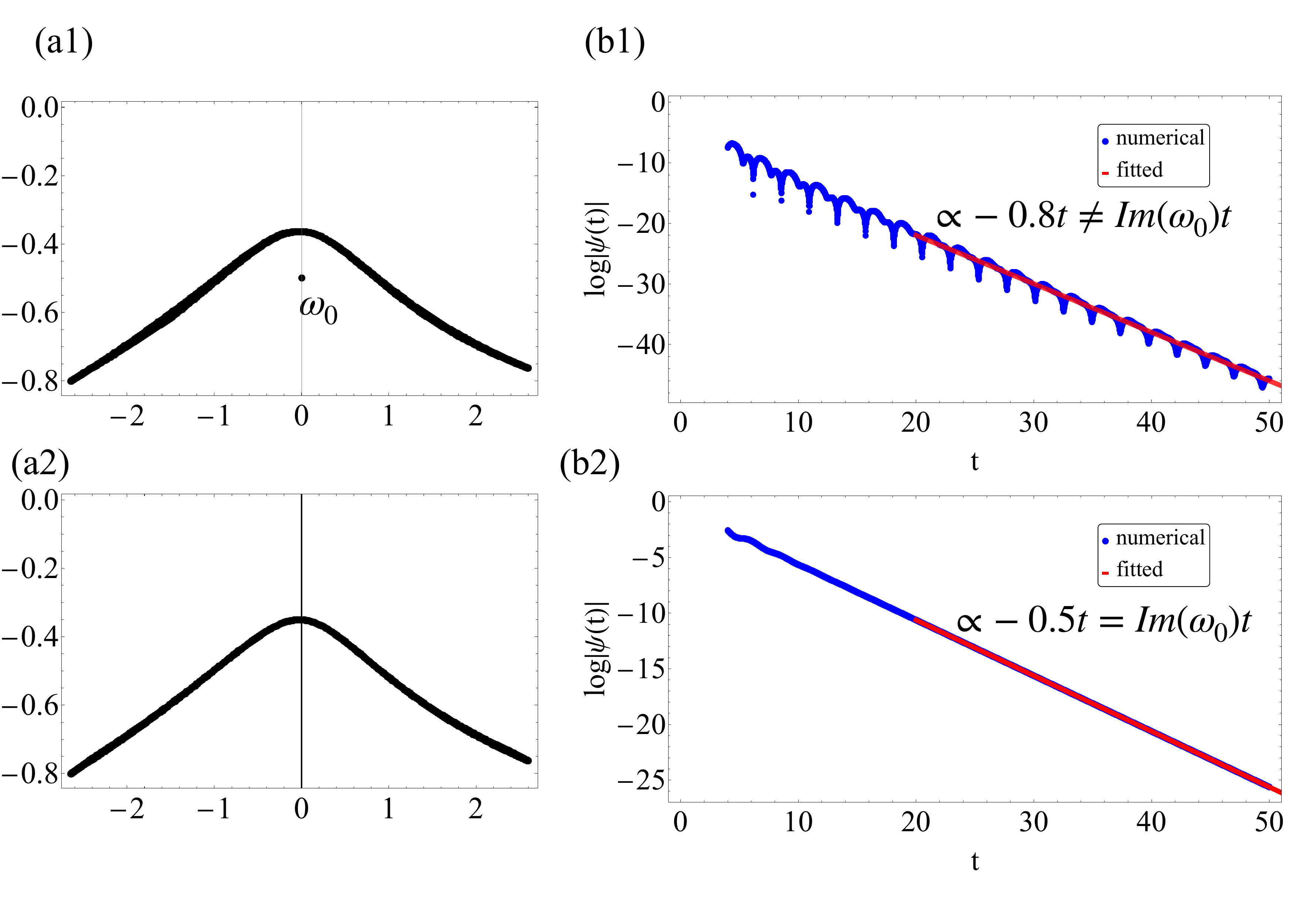}
    \caption{
		(a1) Perturbed OBC spectrum of the model in Fig.~\ref{fig:f3} with impurity strength $\lambda_1=g_0(\omega_0)^{-1}\approx -0.11-2.37i$, where $\omega_0=-0.5i$. Thus, there is a bound state at $\omega_0$ (marked in the figure).
		(b1) $\log|\bra{0}e^{-i\hat{H}t}\ket{0}|$ for $\lambda_1$. The dynamics is still dominated by the branch-cut contribution ($\omega_1$ in Fig.~\ref{fig:f3}) even when a bound state with larger imaginary part exists. (a2), (b2): similar to (a1), (b1) but with impurity strength $\lambda_2=\tilde{g}_0(\omega_0)^{-1}\approx -0.017+1.221i$. Here there is no bound state, but a DP at $\omega_0$ is excited. Thus, the dynamics is dominated by the DP at $\omega_0$.
	}
    \label{fig:f4}
\end{figure}

Figure~\ref{fig:f4} illustrates both consequences for the same OBC model with two different impurity strengths, using the bulk survival amplitude $\langle 0|e^{-i\hat H t}|0\rangle$.
In Fig.~\ref{fig:f4}(a1,b1), the impurity is chosen so that a bound state exists at $\omega_0$, yet the long-time decay is governed by the continuum-edge contribution $e^{-i\omega_1 t}$ rather than by $e^{-i\omega_0 t}$, even though $\Im\omega_0>\Im\omega_1$; the bound state is spectrally present but dynamically dark.
In Fig.~\ref{fig:f4}(a2,b2), the impurity is chosen so that no bound state exists, while a DP at $\omega_0$ dominates the late-time dynamics.

\emph{{\clr Conclusion and Discussion.}---}~
We have shown that late-time impurity scattering in non-Hermitian lattices is organized not by the static spectrum alone, but by the analytic continuation of the Green's function relevant to real-time dynamics. 
The resulting dynamical poles need not coincide with static impurity bound states, which leads to a two-sided static-dynamical mismatch: a DP may appear without any bound-state counterpart, while a static bound state may be dynamically dark. 
At the same time, complex branch points of the continued Green's function act as non-Hermitian continuum edges and generate the incoherent asymptotic background.

This framework naturally recovers the Hermitian limit, where DPs have two fates: those on the real axis coincide with ordinary bound states, while those in the lower half-plane correspond to resonances, i.e., Gamov states~\cite{Madrid2002,Correia2025,Wolkanowski2013,Yapa2023}.
In the Hermitian case, any resonance is always subleading, and every bound state has a DP counterpart and thus remains dynamically visible, recovering the standard result wherein each late-time exponential is the fingerprint of a bound state.
Non-Hermiticity lifts this constraint: DPs can dominate the late-time behavior and bound states can become dynamically dark.

This picture extends beyond the rank-one impurity studied here. 
For a compactly supported impurity potential $\hat V$, the scalar condition $1-\lambda\tilde g_0(\omega)=0$ is replaced by the pole condition of the analytically continued $T$ matrix, i.e., $\det[1-\hat V \tilde G_0(\omega)]=0$, where $\tilde G_0(\omega)$ is the analytically continued Green's function.  
Its matrix elements $\bra{x}\tilde G_0(\omega)\ket{y}$ can be constructed by the same root-selection logic used for $\tilde g_0(\omega)$.
Dynamical poles are therefore not peculiar to the simplest defect model, but are the discrete poles of the time-relevant continuation that control late-time exponentials.

The same analytic-continuation viewpoint also applies to unperturbed dynamics. 
For an initial bulk excitation, the long-time evolution of $\bra{0}e^{-i\hat H_0 t}\ket{0}$ is controlled entirely by branch points and reproduces the relevant saddle-point description of edge dynamics~\cite{Fang2025,Xue2025}. 
As shown in the Supplementary Material~\cite{SupMat}, the branch-point condition is equivalent to the RSP condition, so the branch-point construction also provides an algorithmic way to locate the corresponding saddles.

With an impurity present, however, not only the pole sector but also the continuum-edge sector is reorganized nonperturbatively. 
In one dimension, the unperturbed branch-point singularity gives $\tilde g_0(\omega)\sim(\omega-\omega_b)^{-1/2}$, while the impurity-dressed propagator acquires the factor $\tilde g_0(\omega)[1-\lambda\tilde g_0(\omega)]^{-1}\sim -\lambda^{-1} - \lambda^{-2}(\omega-\omega_b)^{1/2}$ near $\omega_b$\footnote{Observe that $\frac{z^{-1/2}}{1-\lambda z^{-1/2}}=-\frac{1}{\lambda}\frac{1}{1-\lambda^{-1}z^{1/2}}\sim -\lambda^{-1}(1+\lambda^{-1}z^{1/2})$}. 
Consequently, the continuum-edge contribution is suppressed from the unperturbed $t^{-1/2}e^{-i\omega_b t}$ behavior~\cite{Fang2025,Xue2025,SupMat} to the dressed $t^{-3/2}e^{-i\omega_b t}$ behavior. 
Thus the impurity does not merely create dynamical poles; it also reshapes the non-Hermitian continuum-edge response in a genuinely nonperturbative way.

\bibliographystyle{apsrev4-2}
\bibliography{Refs_MainText}

\newpage 
\appendix
\setcounter{equation}{0}  
\setcounter{figure}{0}  
\renewcommand{\thefigure}{A\arabic{figure}}
\renewcommand{\theequation}{A\arabic{equation}}
\begin{widetext}

{\begin{center}
		{\bf \large Supplemental Material: Dynamical Poles in Non-Hermitian Impurity Scattering }
\end{center}}

\section{Appendix I: Step-by-step Derivation of Scattering in Hatano-Nelson model}
\subsection{Model set}
Recall the Hatano-Nelson model under PBC in the maintext, 
\begin{equation}
	\begin{split}
		\hat{H}_0 &= \sum_{k\in \text{BZ}} (e^{ik} + J e^{-ik} - i\kappa)\ket{k}\bra{k} \\
        &= \sum_{k\in \text{BZ}} (1+J)\cos k + i\big((1-J)\sin k-\kappa\big)\ket{k}\bra{k},
        \label{eq:H0}
	\end{split}
\end{equation}
where $J\in (0,1)$ is the reduced hopping and $\kappa >1-J $ is the overall damping moving the spectrum below real axis.

From orthogonal decomposition, 
\begin{equation}
    \begin{split}
        g_0(\omega) :&= \bra{0}(\omega-\hat{H}_0)^{-1}\ket{0}  \\
        &=\frac{1}{2\pi} \int_0^{2\pi} dk \frac{1}{\omega - (e^{ik} + J e^{-ik} - i\kappa)},
    \end{split}
\end{equation}
and by changing variable $z=e^{ik}$ and applying residue theorem,
\begin{equation}
    \begin{split}
        g_0(\omega)&=  \frac{1}{2\pi i} \oint_{|z|=1} dz \frac{1}{(\omega+i \kappa)z -z^2 -J} \\
        &= \begin{cases}
			((\omega+i\kappa)^2 - 4 J)^{-1/2} &\,,\, \omega \in \Omega_0, \\
			0 &\,,\, \omega \in \Omega_1,
        \end{cases} \label{eq:g0}
    \end{split}
\end{equation}
as in the maintext. Note that $g_0(\omega)|_{\Omega_1}=0$ since both roots,$z_{\pm}= \frac{(\omega+i\kappa)\pm\sqrt{(\omega+i\kappa)^2-4J}}{2}$,lies inside the unit circle. And one can check for $\omega\in \mathbb{R}$, $g_0(\omega)$ is unchanged when replacing BZ($|z|=1$) by GBZ($|z|=\sqrt{J}$).  
As a sanity check, from Eq.~\eqref{eq:g0}, $g_0(\omega) \to \omega^{-1}$ as $|\omega| \to \infty $. 

For concrete and reference, let's choose the parameter in Fig 1 in the maintext, i.e., $J=1/2, \kappa =1, \lambda = \tilde g_0(\omega_d)^{-1}\approx 1.436i$ with $\omega_d =-0.75 i$. There is no solution for $1-\lambda g_0(\omega)=0$, so no bound state is excited, and the continuum spectrum is unchanged by the impurity.

For reference, we rewrote the integral to be evaluated here,

\begin{equation}
    \psi_k(x\equiv 0,t) := \bra{0}e^{-i\hat{H}t}\ket{k} = -\bra{0}\ket{k} \frac{1}{2\pi i}\int_{-\infty}^\infty  d\omega \frac{1}{1-\lambda g_0(\omega)}\frac{1}{\omega - E_k} e^{-i\omega t} .\label{eq:psik0t} 
\end{equation}
As mentioned in the maintext, $g_0(\omega)$ in Eq.~\eqref{eq:psik0t} can be replaced by its analytic continuation
\begin{equation}
    \begin{split}
    \tilde g_0(\omega) := ((\omega+i\kappa)^2 - 4 J)^{-1/2} , \omega \in \mathbb{C}\setminus(\Gamma_+ \cup \Gamma_- ),
	\end{split} \label{eq:tildeg0}
\end{equation}
with branch points $\omega_{b,\pm} = \pm 2\sqrt{J}-i\kappa$ and branch cut $\Gamma_\pm$ going downward vertically from $\omega_{b,\pm}$ to $-i \infty$.

\subsection{Contour decomposition and evaluation}
Define 
\begin{equation}
    \begin{split}
        F(\omega) := \frac{e^{-i\omega t}}{(\omega - E_k)(1-\lambda \tilde g_0(\omega))}. \label{eq:fomega}
    \end{split}
\end{equation}
Closing the contour downward gives the exact decomposition 
\begin{equation}
    \psi_k(t) = \bra{0}\ket{k} [\Res_{\omega=E_k}F(\omega)+ \sum_{\omega_d\in \tilde S} \Res_{\omega=\omega_d}F(\omega)+ \sum_{s=\pm}\mathcal{I}_s(t)]
\end{equation}
where $\tilde S:=\{\omega: 1- \lambda \tilde g_0(\omega)=0\}$ and $\mathcal{I}_s(t)$ is the keyhole (branch-cut) contribution around $\Gamma_s$

\textbf{(a) incident pole at $\omega = E_k$}
\begin{equation}
    \begin{split}
        \Res_{\omega = E_k}F(\omega) = \frac{e^{-iE_k t}}{1-\lambda \tilde g_0(E_k)} \label{eq:incident}
    \end{split} 
\end{equation}

\textbf{(b) DP pole at $\omega_d \in \tilde S$}

\begin{equation}
    \begin{split}
        \Res_{\omega=\omega_d}F(\omega) = \frac{e^{-i\omega_d t}}{\omega_d -E_k} \cdot  \frac{-1}{\lambda \tilde g_0'(\omega_d)}, \label{eq:DPS}
    \end{split}
\end{equation}
and for Hatano-Nelson model in Eq.~\eqref{eq:H0},
\begin{equation}
    \begin{split}
        \tilde g_0'(\omega) = -\frac{\omega+ i\kappa}{((\omega+i\kappa)^2-4J)^{3/2}}.
    \end{split}
\end{equation}

\textbf{(c) branch cut $\Gamma_s, s=\pm$}

Parametrize the branch cut by 
\begin{equation}
    \omega = \omega_{b,s} - i y, y\in (0,\infty),
\end{equation}
and for simplicity, denote $\Delta_s(y)= \sqrt{(\omega_{b,s}+i\kappa-iy)^2 - 4J}$, then the keyhole evaluation geives the cut contribution
\begin{equation}
    \begin{split}
        \mathcal{I}_s(t) = \frac{1}{2\pi} e^{-i\omega_{b,s}t}\int_0^{\infty}dy \; e^{-yt} \frac{1}{\omega_{b,s}-iy-E_k}\frac{2\lambda\Delta_s(y)}{\Delta_s(y)^2-\lambda^2}. \label{eq:Is}
    \end{split}
\end{equation}
From Watson's lemma~\cite{Miller2006}, the integration in~\eqref{eq:Is} gives factor $t^{-3/2}$. More concretely, since 
\begin{equation}
    \begin{split}
     \frac{2\lambda\Delta_s(y)}{\Delta_s(y)^2-\lambda^2} &= -\frac{4J^{1/4}}{\lambda} e^{-is\pi/4}y^{1/2}(1+O(y)), y\to 0^+, \\
   (\omega_{b,s}-iy-E_k)^{-1} &= (\omega_{b,s}-E_k)^{-1}+O(y) , y\to 0^+
    \end{split}
\end{equation}
and from the identity,
\begin{equation}
    \begin{split}
        \int_0^{\infty}dy \; y^{1/2} e^{-yt} = \frac{\sqrt{\pi}}{2}t^{-3/2},
    \end{split}
\end{equation} 
we have 
\begin{equation}
    \sum_{s=\pm} \mathcal{I}_s \sim \sum_{s=\pm} (-\frac{J^{1/4}e^{-is\pi/4}}{\lambda (\omega_{b,s}-E_k)\sqrt{\pi}})t^{-3/2}e^{-i\omega_{b,s}t}. \label{eq:Issim}
\end{equation}
Putting Eq.~\eqref{eq:incident}\eqref{eq:DPS}\eqref{eq:Issim} together, we have the decomposition in the maintext, 
\begin{equation}
    \psi_k(t) \sim O(e^{-iE_kt}) + O(e^{-i\omega_d t}) + \sum_{s=\pm}O(t^{-3/2}e^{-i\omega_{b,s} t}).
\end{equation}

\subsection{Other matrix elements}
The worked example above evaluates $\psi_k(t)=\bra{0}e^{-i\hat{H}t}\ket{k}$. Here we record the corresponding formulas for the other two classes of matrix elements used in the main text, namely $\bra{0}e^{-i\hat{H}t}\ket{0}$ and $\bra{k'}e^{-i\hat{H}t}\ket{k}$, which correspond to $M_{00}(t)$ and $M_{k'k}(t)$ there. Since $g_0(\omega)=\tilde g_0(\omega)$ on the real axis, the exact real-axis integrals can be written directly in terms of $\tilde g_0$ before contour deformation.

\textbf{(a) local survival amplitude $\bra{0}e^{-i\hat{H}t}\ket{0}$}

From $\hat{G}=\hat{G}_0+\hat{G}_0\hat{T}\hat{G}_0$ with $\hat{T}(\omega)=\lambda\ket{0}\bra{0}/(1-\lambda g_0(\omega))$, one finds
\begin{equation}
    \begin{split}
        \bra{0}(\omega-\hat{H})^{-1}\ket{0}
        &= g_0(\omega)+g_0(\omega)\frac{\lambda}{1-\lambda g_0(\omega)}g_0(\omega) \\
        &= \frac{g_0(\omega)}{1-\lambda g_0(\omega)}.
    \end{split}
\end{equation}
Therefore,
\begin{equation}
    \begin{split}
        \bra{0}e^{-i\hat{H}t}\ket{0}
        =-\frac{1}{2\pi i}\int_{-\infty}^{\infty} d\omega \;
        e^{-i\omega t}\frac{\tilde g_0(\omega)}{1-\lambda \tilde g_0(\omega)} .
    \end{split}
\end{equation}
Define
\begin{equation}
    \begin{split}
        F_{00}(\omega):=
        e^{-i\omega t}\frac{\tilde g_0(\omega)}{1-\lambda \tilde g_0(\omega)} .
    \end{split}
\end{equation}
Closing the contour downward gives the exact decomposition
\begin{equation}
    \begin{split}
        \bra{0}e^{-i\hat{H}t}\ket{0}
        = \sum_{\omega_d\in \tilde S}\Res_{\omega=\omega_d}F_{00}(\omega)
        + \sum_{s=\pm}\mathcal{I}^{00}_s(t).
    \end{split}
\end{equation}
There is no kinematic pole in this channel, because neither external state produces a factor $(\omega-E_k)^{-1}$.

For a DP $\omega_d\in\tilde S$,
\begin{equation}
    \begin{split}
        \Res_{\omega=\omega_d}F_{00}(\omega)
        = -\frac{e^{-i\omega_d t}}{\lambda^2\tilde g_0'(\omega_d)}.
    \end{split}
\end{equation}
Using the same parametrization of the branch cuts as in I.B,
$\omega=\omega_{b,s}-iy$ with $y>0$ and
$\Delta_s(y)=\sqrt{(\omega_{b,s}+i\kappa-iy)^2-4J}$, the keyhole contribution is
\begin{equation}
    \begin{split}
        \mathcal{I}^{00}_s(t)
        = \frac{1}{2\pi}e^{-i\omega_{b,s}t}
        \int_0^\infty dy \; e^{-yt}\,
        \frac{2\Delta_s(y)}{\Delta_s(y)^2-\lambda^2}.
    \end{split}
\end{equation}
Near the branch point,
\begin{equation}
    \begin{split}
        \Delta_s(y)
        &= 2J^{1/4}e^{-is\pi/4}y^{1/2}(1+O(y)), \\
        \frac{2\Delta_s(y)}{\Delta_s(y)^2-\lambda^2}
        &= -\frac{4J^{1/4}}{\lambda^2}e^{-is\pi/4}y^{1/2}(1+O(y)),
        \qquad y\to 0^+ .
    \end{split}
\end{equation}
Applying Watson's lemma then gives
\begin{equation}
    \begin{split}
        \mathcal{I}^{00}_s(t)
        \sim
        -\frac{J^{1/4}e^{-is\pi/4}}{\lambda^2\sqrt{\pi}}\,
        t^{-3/2}e^{-i\omega_{b,s}t}.
    \end{split}
\end{equation}
Hence
\begin{equation}
    \begin{split}
        \bra{0}e^{-i\hat{H}t}\ket{0}
        \sim
        \sum_{\omega_d\in\tilde S} O(e^{-i\omega_d t})
        + \sum_{s=\pm} O(t^{-3/2}e^{-i\omega_{b,s}t}),
    \end{split}
\end{equation}
so the long-time dynamics is controlled only by DPs and branch points.

\textbf{(b) momentum-resolved matrix element $\bra{k'}e^{-i\hat{H}t}\ket{k}$}

From the same resolvent formula,
\begin{equation}
    \begin{split}
        \bra{k'}(\omega-\hat{H})^{-1}\ket{k}
        = \frac{\delta_{k'k}}{\omega-E_k}
        + \braket{k'}{0}\braket{0}{k}\,
        \frac{\lambda}{(\omega-E_k)(\omega-E_{k'})(1-\lambda g_0(\omega))}.
    \end{split}
\end{equation}
Therefore,
\begin{equation}
    \begin{split}
        \bra{k'}e^{-i\hat{H}t}\ket{k}
        = \delta_{k'k}e^{-iE_k t}
        - \braket{k'}{0}\braket{0}{k}\,
        \frac{\lambda}{2\pi i}\int_{-\infty}^{\infty} d\omega \;
        \frac{e^{-i\omega t}}{(\omega-E_k)(\omega-E_{k'})(1-\lambda \tilde g_0(\omega))}.
    \end{split}
\end{equation}
In the following we assume $E_k\neq E_{k'}$; the cases $k'=k$ or accidental degeneracy lead to a double pole and should be treated separately.

Define
\begin{equation}
    \begin{split}
        F_{k'k}(\omega):=
        \frac{e^{-i\omega t}}{(\omega-E_k)(\omega-E_{k'})(1-\lambda \tilde g_0(\omega))}.
    \end{split}
\end{equation}
Closing the contour downward yields
\begin{equation}
    \begin{split}
        \bra{k'}e^{-i\hat{H}t}\ket{k}
        = \delta_{k'k}e^{-iE_k t}
        + \braket{k'}{0}\braket{0}{k}
        \left[
            \mathcal{R}^{k'k}_{k}
            + \mathcal{R}^{k'k}_{k'}
            + \sum_{\omega_d\in\tilde S}\mathcal{R}^{k'k}_{\omega_d}
            + \sum_{s=\pm}\mathcal{I}^{k'k}_s(t)
        \right],
    \end{split}
\end{equation}
where the two kinematic poles contribute
\begin{equation}
    \begin{split}
        \mathcal{R}^{k'k}_{k}
        &:= \lambda \Res_{\omega=E_k}F_{k'k}(\omega)
        = \frac{\lambda e^{-iE_k t}}{(E_k-E_{k'})(1-\lambda \tilde g_0(E_k))}, \\
        \mathcal{R}^{k'k}_{k'}
        &:= \lambda \Res_{\omega=E_{k'}}F_{k'k}(\omega)
        = \frac{\lambda e^{-iE_{k'} t}}{(E_{k'}-E_k)(1-\lambda \tilde g_0(E_{k'}))},
    \end{split}
\end{equation}
and each DP gives
\begin{equation}
    \begin{split}
        \mathcal{R}^{k'k}_{\omega_d}
        &:= \lambda \Res_{\omega=\omega_d}F_{k'k}(\omega) \\
        &= -\frac{e^{-i\omega_d t}}{(\omega_d-E_k)(\omega_d-E_{k'})\tilde g_0'(\omega_d)} .
    \end{split}
\end{equation}
The branch-cut contribution is
\begin{equation}
    \begin{split}
        \mathcal{I}^{k'k}_s(t)
        = \frac{\lambda}{2\pi}e^{-i\omega_{b,s}t}
        \int_0^\infty dy \; e^{-yt}
        \frac{1}{(\omega_{b,s}-iy-E_k)(\omega_{b,s}-iy-E_{k'})}
        \frac{2\lambda\Delta_s(y)}{\Delta_s(y)^2-\lambda^2}.
    \end{split}
\end{equation}
Using the same near-endpoint expansion as in I.B,
\begin{equation}
    \begin{split}
        \mathcal{I}^{k'k}_s(t)
        \sim
        -\frac{J^{1/4}e^{-is\pi/4}}{(\omega_{b,s}-E_k)(\omega_{b,s}-E_{k'})\sqrt{\pi}}\,
        t^{-3/2}e^{-i\omega_{b,s}t}.
    \end{split}
\end{equation}
Thus
\begin{equation}
    \begin{split}
        \bra{k'}e^{-i\hat{H}t}\ket{k}
        \sim
        \delta_{k'k}e^{-iE_k t}
        + O(e^{-iE_k t})
        + O(e^{-iE_{k'} t})
        + \sum_{\omega_d\in\tilde S} O(e^{-i\omega_d t})
        + \sum_{s=\pm} O(t^{-3/2}e^{-i\omega_{b,s}t}).
    \end{split}
\end{equation}
In other words, this channel contains the free propagation term, two additional kinematic poles tied to the incident and detected plane waves, and then the same DP and branch-point contributions that organize the late-time dynamics in the other matrix elements.

\section{Appendix II: Concrete construction of $\tilde g_0(\omega)$ of model with next-nearest-neighbor hopping}
In this section, we demonstrate the idea of analytic continuation of Green's function $g_0(\omega)$ with next-nearest-neighbor hopping model.
\subsection{Model set}\label{secIIA}
\textbf{(a) Model and contour representation.}

We consider a one-dimensional translationally invariant tight-binding model with finite range 2 hopping, whose Bloch Hamiltonian is the Laurent polynomial
\begin{equation}
    \begin{split}
        h(z)= \sum_{n=-2}^2 t_n z^n , z=e^{ik},t_n\in \mathbb{C}.    
    \end{split}
\end{equation}
Here, we assume the spectrum lies below the real axis(as in the maintext), e.g., by a uniform shift $\hat{H}\to \hat{H}-i\kappa \hat{\mathbb{I}}$.

Using the Bloch decomposition, the Green's function admits the standard contour representation
\begin{equation}
    \begin{split}
        g_0(\omega):&= \bra{0}(\omega-\hat{H}_0)^{-1}\ket{0} = \oint_{|z|=1} \frac{dz}{2\pi i }\frac{z}{Q(z;\omega)}, \\
        Q(z;\omega) :&= z^2(\omega-h(z)), 
    \end{split}
\end{equation}
and it follows that $Q(z;\omega)$ is a quartic polynomial in $z$ for fixed $\omega$. 
\\

\textbf{(b) Root structure for $\Im \omega \ge 0$.}

Let $\{z_j(\omega)\}$be the four roots (counted with multiplicity) of $Q(z;\omega)$, equivalently $\omega = h(z_j(\omega))$:
\begin{equation}
    \begin{split}
        Q(z_j(\omega),\omega) = 0, j=1,2,3,4.
    \end{split}
\end{equation}
There are no roots on the unit circle. Since $\Im \omega \ge 0$ and the spectrum of $\hat{H}_0$ lies below the real axis, $\omega \neq h(e^{ik})$ for all real $k$. Equivalently, $\omega - h(z)\neq 0, \forall |z| =1$.

Importantly, there are exactly two roots inside and two outside the unit circle. Specifically, for $|\omega|\to \infty$,the dominant terms in $Q(z;\omega)$
 imply two roots scale as $z\sim \omega^{-1/2}$
 (small magnitude) and two scale as $z\sim \omega^{1/2}$
 (large magnitude). Therefore, at a reference point 
$\omega_{\text{ref}}= i R$ with $R\gg 0$ in the upper half-plane, there are exactly two roots inside 
$|z|=1$and two outside. Since for $\Im \omega \ge 0$ no root crosses the unit circle (as shown above), the number of roots inside $|z|=1$ is invariant throughout the closed upper half-plane. Consequently, for all $\Im \omega \ge 0$, exactly two of roots $z_j(\omega)$ satisfy $|z_j(\omega)|<1$ and other two satisfy $|z_j(\omega)|>1$. 
\\

\textbf{(c) Residue evaluation.}

By residue theorem, 
\begin{equation}
    \begin{split}
        g_0(\omega) = \sum_{|z_j(\omega)|<1}\Res_{z=z_j(\omega)} \frac{z}{Q(z;\omega)}.
    \end{split}
\end{equation}
To make the residue structure explicit, factor the quartic polynomial as
\begin{equation}
    \begin{split}
        Q(z;\omega) = -t_2 \prod_{i=1}^4 (z-z_i(\omega)).
    \end{split}
\end{equation}
And for a simple root $z=z_j(\omega)$,the residue is 
\begin{equation}
    \begin{split}
        \Res_{z=z_j(\omega)} \frac{z}{Q(z;\omega)} = -\frac{1}{z_j\,h'(z_j)} = \frac{1}{-t_2} \frac{z_j}{\prod_{i\neq j}(z_j(\omega)-z_i(\omega))}
    \end{split}
\end{equation}
Therefore for $\Im\omega\ge 0$,
\begin{equation}
g_0(\omega)=-\sum_{|z_j(\omega)|<1}\frac{1}{z_j(\omega)\,h'(z_j(\omega))},
\qquad (\Im\omega\ge 0), \label{eq:g0w}
\end{equation}

\subsection{Analytic continuation}

For $\Im \omega \ge 0$, $g_0(\omega)$ is given by the residue sum~\eqref{eq:g0w}.
We now construct an analytic continuation $\tilde g_0(\omega)$ to the lower half-plane
(and across point gaps) by fixing a choice of root branches and continuing them as analytic functions of $\omega$.

\textbf{(a) Ramification set in the $\omega$-plane.}

By the complex Implicit Function Theorem, if $(z_0,\omega_0)$ satisfies $Q(z_0,\omega_0)=0$ but $ \partial_z Q(z_0;\omega_0)\neq 0$ there exists a neighborhood $U\ni \omega_0$ and a unique analytic function  $z(\omega)$ on $U$ such that $z(\omega_0)=z_0$ and  $Q(z(\omega),\omega)=0 $ for all $\omega \in U $. Hence, each simple root $z_j(\omega)$ is analytic in $\omega$ locally away from the set where $Q$ has multiple roots.

Equivalently, ramification points of the algebraic function $z(\omega)$ arise when $Q$ has a multiple root, i.e.
\begin{equation}
Q(z;\omega)=0, \qquad \partial_z Q(z;\omega)=0.
\end{equation}
Since $Q(z;\omega)=z^2(\omega-h(z))$, these two conditions are equivalent
to
\begin{equation}
\omega = h(z), \qquad h'(z)=0.
\label{eq:branchpoint_condition}
\end{equation}
We denote the corresponding ramification set by
\begin{equation}
    \mathcal{R}:= \{\omega : \exists z\neq 0, s.t., Q(z,\omega) = \partial_z Q(z;\omega) =0 \} = \{h(z_b):z_b\neq 0 , h'(z_b)=0\}.
\end{equation}
We exclude $z=0$ because the equivalence to is obtained after dividing by powers of $z$. In the generic range-2 case $t_{\pm 2}\neq 0, Q(z=0,\omega)\neq 0,\forall \omega$ . Otherwise, the degree is reduced and the analysis adapts verbatim with lower degree.

For finite-range hopping, $h'(z)=0$ is a finite-degree polynomial equation, hence there are finitely many ramification point $\omega_b$ for $z(\omega)$. But notably, $\omega_b$ is not necessarily a ramification point of $\tilde g_0(\omega)$ due to possible cancellation below.

As a computational remark, $\mathcal{R}$ is the zero set of the discriminant $\Delta(\omega)$ of $Q(\cdot;\omega )$,i.e., $\Delta(\omega)=0$ where $\Delta(\omega):= \Res_z(Q,\partial_z Q)$.
\\

\textbf{(b) Fixing the sheet: definition of $\tilde g_0(\omega)$.}

Choose a reference point $\omega_{\rm ref}=iR$ with $R\gg 1$.
As shown in Sec.~\ref{secIIA}, at $\omega_{\rm ref}$ there are exactly two roots inside the unit circle and two outside.
Label the two roots inside $|z|=1$ at $\omega_{\rm ref}$ as $z_1(\omega_{\rm ref}),z_2(\omega_{\rm ref})$. We define $\tilde g_0(\omega)$ by 
\begin{equation}
\tilde g_0(\omega):=-\sum_{j=1}^2 \frac{1}{z_j(\omega)\,h'(z_j(\omega))},
\label{eq:g0_tilde_def}
\end{equation}
where $z_{1,2}(\omega)$ are to be understood as the continuations of $z_{1,2}(\omega_{\rm ref})$ specified below. 
\\

\textbf{(c) Branch point and cut of $\tilde g_0(\omega)$.}

Analytic continuation of $z_i(\omega)$ around a point $\omega_b \in \mathcal{R}$ permutes the four root labels $\{z_1,z_2,z_3,z_4\}$ (monodromy). However, $\tilde g_0(\omega)$ in Eq.~\eqref{eq:g0_tilde_def} depends on $z_1,z_2$ only through a symmetric sum.
Therefore, a permutation $z_1 \leftrightarrow z_2$ leaves $\tilde g_0(\omega)$ invariant, and likewise a permutation $z_3 \leftrightarrow z_4$ is irrelevant.
Consequently, $\tilde g_0$ can fail to be single-valued near $\omega_b$ only if the analytic continuation exchanges one of the chosen branches
$\{z_1,z_2\}$ with one of the complementary branches
$\{z_3,z_4\}$.

We denote by $B\subset \mathcal{R}$ the subset of multiple-root energies for which this mixing of the two sets of branches occurs.

It follows that $\tilde g_0(\omega)$ can have a singularity at $\omega_b$ only if $\omega_b \in B$. For generic double-root energies $\omega_b \in \mathcal{R}$, the local square-root analysis below shows that $\omega_b \in B$ indeed produces a square-root branch singularity in $\tilde g_0(\omega)$.

To make $\tilde g_0$ single-valued, we remove from the $\omega$-plane a set of cuts $\Gamma$ that emanate from the points $\omega_b\in B$ and extend to infinity.
For definiteness (and for later inverse Fourier/Laplace contour deformations), we choose fences:
\begin{equation}
\Gamma := \bigcup_{\omega_b\in B}\left\{\omega=\omega_b-i y : y\in[0,\infty)\right\}.
\label{eq:b:fences}
\end{equation}
Let $\Omega:=\mathbb C\setminus\Gamma$. On $\Omega$, $\tilde g_0(\omega)$ is single-valued and analytic. Moreover, the boundary condition is satisfied,
\begin{equation}
\tilde g_0(\omega)=g_0(\omega),\qquad \Im\omega>0 .
\label{eq:b:principal_sheet_bc}
\end{equation}
\\

\textbf{(d) Local square-root structure near a ramification point.}

Let $\omega_b=h(z_b)\in \mathcal{R}$ with $h'(z_b)=0$ and $h''(z_b)\neq 0$.
Expanding $h(z)$ near $z=z_b$ gives
\begin{equation}
h(z)=\omega_b+\frac{1}{2}h''(z_b)(z-z_b)^2+O\!\left((z-z_b)^3\right).
\end{equation}
Hence the two roots that collide at $\omega_b$ behave as
\begin{equation}
z_\pm(\omega)=z_b \pm \sqrt{\frac{2(\omega-\omega_b)}{h''(z_b)}} + O(\omega-\omega_b),
\end{equation}
which shows the square-root branching of $z(\omega)$ at $\omega_b$.
Moreover, since $h'(z)=h''(z_b)(z-z_b)+\cdots$, we obtain
\begin{equation}
\frac{1}{z_\pm(\omega)\,h'(z_\pm(\omega))}
=
\pm \frac{1}{z_b}\frac{1}{\sqrt{2h''(z_b)}}\frac{1}{\sqrt{\omega-\omega_b}}
+O(1).
\label{eq:root_singularity}
\end{equation}

Equation Eq.~\eqref{eq:root_singularity} confirms that $\tilde g_0(\omega)$ has a branch singularity at $\omega_b$
iff exactly one of the two colliding branches is included in the sum Eq.~\eqref{eq:g0_tilde_def}.
If both colliding branches are included (or both are excluded), the leading $\pm 1/\sqrt{\omega-\omega_b}$ terms cancel.
Equivalently, $\omega_b$ is a branch point of $\tilde g_0$ precisely when the monodromy around $\omega_b$
exchanges one root from $\{z_1,z_2\}$ with one root from $\{z_3,z_4\}$ (the complementary set at $\omega_{\rm ref}$). Thus, $\omega_b\in\mathcal{R}$ becomes a branch point of $\tilde g_0(\omega)$ iff $\omega_b \in B$.
\\

\subsection{Branch-cut contribution and universal continuum-edge tails}
The main text keeps only the universal statement that branch points in $B$ produce continuum-edge contributions of the form $I_{\omega_b}(t)\sim t^{-3/2}e^{-i\omega_b t}$ in one dimension. Here we record the contour formula behind that result and its extension to the matrix elements used in the paper. Throughout, let $\omega_b\in B$ be a branch point of the analytically continued Green's function, and choose the corresponding branch cut $\Gamma_b$ to run vertically downward from $\omega_b$.

For any matrix element whose frequency-domain representation can be written as
\begin{equation}
\mathcal{A}(t)=-\frac{1}{2\pi i}\int_{-\infty}^{\infty}d\omega\;F(\omega)e^{-i\omega t},\qquad (t>0),
\end{equation}
closing the contour into $\Im\omega<0$ gives pole contributions together with keyhole integrals around the cuts. The contribution from $\Gamma_b$ can be written as
\begin{equation}
I_{\omega_b}(t)=\frac{e^{-i\omega_b t}}{2\pi i}\int_0^{\infty}dy\;e^{-yt}\Big[F(\omega_b-iy+0^+)-F(\omega_b-iy-0^+)\Big].
\label{eq:sm_branchcut_generic}
\end{equation}
Thus the long-time asymptotics is fixed entirely by the discontinuity of $F(\omega)$ across the cut near $y=0^+$. Writing
\begin{equation}
\mathrm{Disc}\,F(\omega_b-iy):=F(\omega_b-iy+0^+)-F(\omega_b-iy-0^+),
\end{equation}
if
\begin{equation}
\mathrm{Disc}\,F(\omega_b-iy)\sim \alpha_b\,y^{\sigma_b},\qquad y\to 0^+,
\label{eq:sm_branchcut_sigma}
\end{equation}
then Watson's lemma~\cite{Miller2006} yields
\begin{equation}
I_{\omega_b}(t)\sim \frac{\alpha_b\,\Gamma(\sigma_b+1)}{2\pi i}\,t^{-\sigma_b-1}e^{-i\omega_b t}.
\label{eq:sm_branchcut_watson}
\end{equation}
More generally, if the discontinuity is logarithmically dressed,
\begin{equation}
\mathrm{Disc}\,F(\omega_b-iy)\sim \beta_b\,\frac{y^{\sigma_b}}{(\log y)^2},\qquad y\to 0^+,
\label{eq:sm_branchcut_sigma_log}
\end{equation}
then setting $s=yt$ in Eq.~\eqref{eq:sm_branchcut_generic} gives
\begin{equation}
I_{\omega_b}(t)\sim \frac{\beta_b\,e^{-i\omega_b t}}{2\pi i\,t^{\sigma_b+1}}
\int_0^\infty ds\;e^{-s}\frac{s^{\sigma_b}}{(\log s-\log t)^2}.
\end{equation}
For fixed $s$ and large $t$, $(\log s-\log t)^{-2}\sim (\log t)^{-2}$, so Watson's lemma again yields
\begin{equation}
I_{\omega_b}(t)\sim \frac{\beta_b\,\Gamma(\sigma_b+1)}{2\pi i}\,
\frac{e^{-i\omega_b t}}{t^{\sigma_b+1}(\log t)^2}.
\label{eq:sm_branchcut_watson_log}
\end{equation}
In particular, if $F(\omega)\sim a_b\log(\omega-\omega_b)$, then $\mathrm{Disc}\,F(\omega_b-iy)\sim 2\pi i\,a_b$, so the clean $d=2$ local propagator gives a $t^{-1}e^{-i\omega_b t}$ branch-cut contribution. The impurity-dressed case in $d=2$, where the discontinuity acquires a $(\log y)^{-2}$ envelope, is analyzed in Sec.~\ref{sec:higher_dim} below. This is the general branch-cut rule used in the main text.

For the one-dimensional impurity problem discussed there, the relevant input is the local square-root singularity of the continued Green's function. Near a multiple-root energy $\omega_b\in B$, the collision of two roots implies
\begin{equation}
\tilde g_0(\omega)\sim c_b\,(\omega-\omega_b)^{-1/2},
\label{eq:sm_tildeg_sqrt}
\end{equation}
with a nonzero coefficient $c_b$. In impurity-dressed amplitudes, this singularity appears through the factor $[1-\lambda\tilde g_0(\omega)]^{-1}$, so that generically
\begin{equation}
\frac{1}{1-\lambda\tilde g_0(\omega)}\sim -\frac{1}{\lambda c_b}(\omega-\omega_b)^{1/2}.
\label{eq:sm_dressed_kernel_sqrt}
\end{equation}
Therefore the discontinuity of the dressed kernel scales as $y^{1/2}$, i.e. $\sigma_b=1/2$, and Eq.~\eqref{eq:sm_branchcut_watson} gives
\begin{equation}
I_{\omega_b}(t)\sim C_{\omega_b}\,t^{-3/2}e^{-i\omega_b t}.
\label{eq:sm_branchcut_t32}
\end{equation}
This is the universal continuum-edge contribution quoted in the main text.

For momentum-resolved scattering, the integrand contains additional factors $(\omega-E_k)^{-1}(\omega-E_{k'})^{-1}$. As long as the branch point does not collide with a kinematic pole, these factors are regular at $\omega_b$ and only modify the channel-dependent coefficient $C_{\omega_b}^{k'k}$, not the exponent. The same conclusion holds for local survival amplitudes such as $\bra{0}e^{-i\hat H t}\ket{0}$: the branch-cut piece again scales as $t^{-3/2}e^{-i\omega_b t}$, with a different prefactor. By contrast, in unperturbed dynamics the kernel is proportional to $\tilde g_0(\omega)$ itself, whose discontinuity scales as $y^{-1/2}$, leading to the familiar $t^{-1/2}e^{-i\omega_b t}$ edge contribution. The impurity therefore changes the continuum-edge response nonperturbatively, from the unperturbed $t^{-1/2}$ law to the dressed $t^{-3/2}$ law, while also generating dynamical poles through the same analytically continued structure.

\section{Appendix III: Local behavior of Green's function in higher dimensions}\label{sec:higher_dim}
In 1D we have a finite polynomial equation $\omega-h(z)=0$ with roots $z_j(\omega)$.
Ramification points are $\omega_b$ where two roots collide, equivalently $h'(z_b)=0$, and the local square-root behavior follows.

In $d\ge2$, the analogous object is the holomorphic map
\begin{equation}
    \begin{split}
        h: \mathbb{C}^d \to \mathbb{C}, \qquad \mathbf{z} \mapsto h(\mathbf{z}),
    \end{split}
\end{equation}
(or $Q(\mathbf{z};\omega)=\det(\omega-H(\mathbf{z}))$ in the multiband case). The natural ramification set is
\begin{equation}
    \begin{split}
        \omega_b=h(\mathbf{z}_b), \qquad \partial_{z_i}h(\mathbf{z}_b)=0,\ \forall i.
    \end{split}
\end{equation}

Unlike 1D, $\omega-h(\mathbf{z})=0$ does not give isolated roots in $\mathbf{z}$; it gives a $d-1$-dimensional hypersurface. Thus, the higher-dimensional analogue of a root collision is the pinching of the integration cycle by that hypersurface at a critical point.

Concretely, assume that the integration region of $g_0(\omega)$, namely the BZ, can be deformed to some $\mathcal{C}\subset \mathbb{C}^d$ such that for $\omega$ in a punctured neighborhood $0<|\omega-\omega_b|<\epsilon$ and $\mathbf{z}\in\mathcal{C}$, one has $\omega-h(\mathbf{z})\neq0$. Then the analytically continued Green's function admits the local representation
\begin{equation}
    \begin{split}
        \tilde g_0(\omega)
        &= \int_\mathcal{C} \frac{d^d\mathbf{z}}{(2\pi i)^d \prod_i z_i}\frac{1}{\omega-h(\mathbf{z})} \\
        &\sim \int_{|\mathbf{u}|<\Gamma}\frac{d^d\mathbf{u}}{\Delta-\frac{1}{2}\mathbf{u}^T A \mathbf{u}}
        =: I(\Delta), \qquad \Delta:=\omega-\omega_b,
    \end{split}
\end{equation}
where the ``$\sim$'' means equality up to terms analytic in $\Delta$ near $0$.
Here $A$ is the nondegenerate quadratic form obtained from the Hessian of $h(\mathbf{z})$ after a local analytic parametrization, and the cutoff $\Gamma$ affects only the regular part.
Therefore, the universal information near $\omega_b$ is carried by the singular part of $I(\Delta)$, which determines the nonanalytic part $\tilde g_{0,\mathrm{sing}}(\omega)$ of $\tilde g_0(\omega)$.

After a nonsingular linear change of variables, together with an overall nonzero prefactor that is analytic at $\Delta=0$, the singular structure is equivalent to the isotropic radial model
\begin{equation}
    \begin{split}
        I_d(\Delta)\sim \int_0^\Gamma \frac{r^{d-1}\,dr}{\Delta-r^2}.
    \end{split}
\end{equation}
The ultraviolet-divergent polynomial pieces generated from this integral contribute only to the analytic part of $\tilde g_0(\omega)$; the branch singularity is fixed by the remaining nonanalytic term.

\subsection{Odd dimensions}
Let $d=2m+1$ with $m\ge1$. Then
\begin{equation}
    \begin{split}
        \frac{r^{2m}}{\Delta-r^2}
        = -\sum_{j=0}^{m-1}\Delta^j r^{2m-2-2j}
        + \frac{\Delta^m}{\Delta-r^2}.
    \end{split}
\end{equation}
After integration over $r$, the finite polynomial terms produce only analytic contributions in $\Delta$, while the last term yields the nonanalytic piece
\begin{equation}
    \begin{split}
        I_{2m+1,\mathrm{sing}}(\Delta)
        \sim \Delta^m \int_0^\Gamma \frac{dr}{\Delta-r^2}
        \sim \Delta^{m-\frac{1}{2}}.
    \end{split}
\end{equation}
Hence, for odd dimensions,
\begin{equation}
    \begin{split}
        \tilde g_{0,\mathrm{sing}}(\omega)\sim (\omega-\omega_b)^{\frac{d}{2}-1},
        \qquad d\ \text{odd}.
    \end{split}
\end{equation}

\subsection{Even dimensions}
Let $d=2m$ with $m\ge2$. Using $s=r^2$, one obtains
\begin{equation}
    \begin{split}
        I_{2m}(\Delta)\sim \frac{1}{2}\int_0^{\Gamma^2}\frac{s^{m-1}\,ds}{\Delta-s}.
    \end{split}
\end{equation}
Now
\begin{equation}
    \begin{split}
        \frac{s^{m-1}}{\Delta-s}
        = -\sum_{j=0}^{m-2}\Delta^j s^{m-2-j}
        + \frac{\Delta^{m-1}}{\Delta-s}.
    \end{split}
\end{equation}
Again, the polynomial terms are analytic in $\Delta$, whereas the last term gives
\begin{equation}
    \begin{split}
        I_{2m,\mathrm{sing}}(\Delta)
        \sim \Delta^{m-1}\int_0^{\Gamma^2}\frac{ds}{\Delta-s}
        \sim \Delta^{m-1}\log\Delta.
    \end{split}
\end{equation}
Therefore, for even dimensions with $d\ge4$,
\begin{equation}
    \begin{split}
        \tilde g_{0,\mathrm{sing}}(\omega)\sim (\omega-\omega_b)^{\frac{d}{2}-1}\log(\omega-\omega_b),
        \qquad d\ \text{even},\ d\ge4.
    \end{split}
\end{equation}

\subsection{$d=2$}
For $d=2$, it is convenient to work directly in polar coordinates, $\mathbf{u}=(r\cos\theta,r\sin\theta)$, so that
\begin{equation}
    \begin{split}
        \frac{1}{2}\mathbf{u}^T A \mathbf{u}=r^2 f(\theta),
    \end{split}
\end{equation}
where $f(\theta)$ is a nonzero complex-valued function. If $f(\theta_0)=0$ for some $\theta_0$, then higher-order terms in the local expansion are required and the singularity is nongeneric. Thus, the singular piece is
\begin{equation}
    \begin{split}
        I(\Delta)\sim \int_0^{2\pi} d\theta \int_0^\Gamma \frac{r\,dr}{\Delta-r^2 f(\theta)}.
    \end{split}
\end{equation}
The radial integral is explicit:
\begin{equation}
    \begin{split}
        \int_0^\Gamma \frac{r\,dr}{\Delta-r^2 f(\theta)}
        = -\frac{1}{2f(\theta)}\log\bigl(\Delta-r^2 f(\theta)\bigr)\Big|_0^\Gamma.
    \end{split}
\end{equation}
Hence
\begin{equation}
    \begin{split}
        I(\Delta)
        = \left(\frac{1}{2}\int_0^{2\pi}\frac{d\theta}{f(\theta)}\right)\log\Delta
        + I_{\mathrm{reg}}(\Delta),
    \end{split}
\end{equation}
where $I_{\mathrm{reg}}(\Delta)$ is analytic in $\Delta$ near $0$. Therefore,
\begin{equation}
    \begin{split}
        \tilde g_{0,\mathrm{sing}}(\omega)\sim \log(\omega-\omega_b),
        \qquad d=2.
    \end{split}
\end{equation}

The odd/even structure above is the same as the Hermitian band-edge singularity in the corresponding dimension \cite{Joyce2003}.
For the clean local propagator
\begin{equation}
    \begin{split}
        A_0(t):=\bra{0}e^{-i\hat H_0 t}\ket{0},
    \end{split}
\end{equation}
the logarithmic singularity in $d=2$ implies that the branch-cut contribution from $\omega_b$ scales as
\begin{equation}
    \begin{split}
        A_{0,\omega_b}(t)\sim t^{-1}e^{-i\omega_b t}.
    \end{split}
\end{equation}
By contrast, once the impurity resummation is included, the relevant kernels for both momentum-resolved scattering and local survival amplitudes take the generic form
\begin{equation}
    \begin{split}
        F(\omega)=R(\omega)\frac{1}{1-\lambda \tilde g_0(\omega)}
        \qquad \text{or} \qquad
        F(\omega)=R(\omega)\frac{\tilde g_0(\omega)}{1-\lambda \tilde g_0(\omega)},
    \end{split}
\end{equation}
where $R(\omega)$ is regular at $\omega_b$. Using the local expansion
\begin{equation}
    \begin{split}
        \tilde g_0(\omega)=a_b\log\Delta+b_b+O(\Delta\log\Delta),\qquad
        \Delta:=\omega-\omega_b,\qquad a_b\neq0,
    \end{split}
\end{equation}
one finds, for $\lambda\neq0$,
\begin{equation}
    \begin{split}
        \frac{1}{1-\lambda \tilde g_0(\omega)}
        \sim -\frac{1}{\lambda a_b\,\log\Delta},
    \end{split}
\end{equation}
and
\begin{equation}
    \begin{split}
        \frac{\tilde g_0(\omega)}{1-\lambda \tilde g_0(\omega)}
        = -\frac{1}{\lambda}+O\!\left(\frac{1}{\log\Delta}\right).
    \end{split}
\end{equation}
The constant term $-1/\lambda$ carries no discontinuity, whereas
\begin{equation}
    \begin{split}
        \mathrm{Disc}\,\frac{1}{\log\Delta}
        = \frac{\log\Delta_- - \log\Delta_+}{\log\Delta_+\log\Delta_-}
        \sim \frac{2\pi i}{(\log y)^2},
        \qquad \Delta=-iy,\quad y\to0^+.
    \end{split}
\end{equation}
Hence in both cases
\begin{equation}
    \begin{split}
        \mathrm{Disc}\,F(\omega_b-iy)\sim \frac{\beta_b}{(\log y)^2},
    \end{split}
\end{equation}
and Eq.~\eqref{eq:sm_branchcut_watson_log} gives the generic impurity-dressed branch-cut contribution
\begin{equation}
    \begin{split}
        I_{\omega_b}(t)\sim \widetilde C_{\omega_b}\,
        \frac{e^{-i\omega_b t}}{t(\log t)^2}.
    \end{split}
\end{equation}

\section{Appendix IV: Relation between branch points and relevant saddle points}

In the main text, we stated that for unperturbed dynamics the branch-point construction reproduces the relevant-saddle-point (RSP) description of non-Hermitian edge dynamics. We now explain this correspondence in the one-dimensional setting relevant to the present work.

For an initial bulk excitation, the local return amplitude of the clean lattice can be written either as a momentum (or $z$) integral,
\begin{equation}
A_0(t):=\langle 0|e^{-i\hat H_0 t}|0\rangle
=
\oint_{|z|=1}\frac{dz}{2\pi i\, z}\,e^{-i h(z)t},
\label{eq:sm_rsp_A0_z}
\end{equation}
or equivalently as a frequency integral,
\begin{equation}
A_0(t)=\frac{1}{2\pi i}\int_{\mathbb{R}+i0^+} d\omega\; e^{-i\omega t}\, g_0(\omega),
\label{eq:sm_rsp_A0_w}
\end{equation}
where
\begin{equation}
g_0(\omega)=\oint_{|z|=1}\frac{dz}{2\pi i\, z}\,\frac{1}{\omega-h(z)}.
\label{eq:sm_rsp_g0}
\end{equation}
The RSP approach analyzes Eq.~\eqref{eq:sm_rsp_A0_z} by deforming the $z$ contour into a sum of Lefschetz thimbles attached to saddle points $z_s$ of $h(z)$, i.e.
\begin{equation}
h'(z_s)=0.
\label{eq:sm_rsp_spcond}
\end{equation}
Our approach instead analyzes Eq.~\eqref{eq:sm_rsp_A0_w} by continuing $g_0(\omega)$ into $\Im\omega<0$ and deforming the $\omega$ contour onto the branch cuts of $\tilde g_0(\omega)$.

These two constructions are equivalent because a saddle point of $h(z)$ is exactly a candidate branch point of $g_0(\omega)$. Indeed, a branch point of $g_0$ occurs when $\omega-h(z)=0$ develops a multiple root:
\begin{equation}
\omega_b-h(z_b)=0,\qquad h'(z_b)=0.
\label{eq:sm_rsp_bpcond}
\end{equation}
Thus the branch-point energy is simply
\begin{equation}
\omega_b=h(z_b),
\label{eq:sm_rsp_es}
\end{equation}
namely the saddle-point energy. Near such a point,
\begin{equation}
\omega-h(z)\simeq (\omega-\omega_b)-\frac{h''(z_b)}{2}(z-z_b)^2,
\label{eq:sm_rsp_local}
\end{equation}
so the two local root branches satisfy
\begin{equation}
z_\pm(\omega)-z_b \propto \sqrt{\omega-\omega_b}.
\label{eq:sm_rsp_sqrt}
\end{equation}
Therefore every generic saddle produces a square-root branching in the $\omega$ plane.

The geometric relation between our branch cut and the Lefschetz thimble is also immediate. Along a descending thimble of the saddle $z_b$, the real part of $h(z)$ is constant while the imaginary part decreases away from the saddle. Since $\omega=h(z)$ on the image of the thimble, its image in the $\omega$ plane is a vertical downward ray
\begin{equation}
\omega=\omega_b-i y,\qquad y>0.
\label{eq:sm_rsp_cut}
\end{equation}
This is precisely the downward branch cut used in our contour deformation. In this sense, the thimble in the $z$ plane and the downward cut in the $\omega$ plane are the same object viewed in two different variables.

What remains is the relevance criterion. In our language, not every multiple root contributes: a branch point contributes only when circling $\omega_b$ exchanges one selected root branch and one unselected root branch. If the monodromy only permutes roots within the selected set, or only within the unselected set, then the symmetric residue sum defining $\tilde g_0(\omega)$ remains analytic and no branch point appears in the continued Green's function. This is exactly the same mechanism that appears in the RSP analysis. There, the relevant energy-space object is also symmetric under exchanges within each root bracket, so such exchanges do not create branching, whereas a cross-bracket exchange does create branching and forces the contour deformation to split at the saddle-point energy. The deformed contour then continues on a single-valued sheet separated by a branch cut starting at that energy.

Therefore, in the present one-dimensional setting,
\begin{equation}
\omega_b\in B
\quad\Longleftrightarrow\quad
z_b \ \text{is an RSP}.
\label{eq:sm_rsp_equiv}
\end{equation}
Equivalently, the set $B$ of branch points of the continued local Green's function is exactly the set of RSP energies. The dominant saddle point (DSP) is then simply the element of $B$ with the largest imaginary part,
\begin{equation}
\omega_{\mathrm{DSP}}=\arg\max_{\omega_b\in B}\Im\omega_b.
\label{eq:sm_rsp_dsp}
\end{equation}
Its contribution is the leading branch-cut term in the frequency-domain formulation.

This equivalence is useful conceptually and algorithmically. Conceptually, it shows that the continuum-edge sector in our formulation is the same asymptotic sector described by the RSP formalism. Algorithmically, it means that one may identify the relevant saddles without explicitly performing a gradient-flow analysis: solve $h'(z)=0$, map each solution to $\omega_b=h(z_b)$, and retain only those for which the local root exchange mixes a selected and an unselected branch. In that way, the branch-point construction provides an alternative route to the RSPs and, in particular, to the DSP.

\end{widetext}

\end{document}